\documentstyle[12pt]{article}

\begin{document}
\begin{titlepage}
\begin{centering}
\vspace{1.5cm}
{\LARGE{\bf Weakly nonlinear investigation of the Saffman-Taylor problem in a rectangular Hele-Shaw cell}}\\
\bigskip\bigskip
Jos\'e A. Miranda\footnote[1]{e-mail:01jamn@npd.ufpe.br} and Michael Widom\footnote[2]{e-mail:widom@andrew.cmu.edu}\\
{\em Department of Physics, Carnegie Mellon University, Pittsburgh, PA 15213}\\
\end{centering}
\begin{abstract}
We analyze the Saffman-Taylor 
viscous fingering problem in rectangular geometry. We investigate 
the onset of nonlinear effects and the basic symmetries of the mode 
coupling equations, highlighting the link between interface asymmetry 
and viscosity contrast. Symmetry breaking occurs through 
enhanced growth of sub-harmonic perturbations. Our results explain 
the absence of finger tip-splitting in the early flow stages, and 
saturation of growth rates compared with the predictions of linear 
stability. 
\end{abstract}
\begin{center}
\end{center}
\hspace{0.7 cm}
\end{titlepage}
\def\carre{\vbox{\hrule\hbox{\vrule\kern 3pt
\vbox{\kern 3pt\kern 3pt}\kern 3pt\vrule}\hrule}}

\baselineskip = 30pt
\section{Introduction}
\label{intro}

Pattern formation occurs in many hydrodynamics settings. 
The Saffman-Taylor problem~\cite{Saf}, in which 
two immiscible viscous fluids move in a narrow space between 
two parallel plates (the so-called Hele-Shaw cell), is a widely 
studied example where a fluid-fluid interface evolves~\cite{Rev}. 
The initially flat interface separating the
two fluids can be destabilized by either a pressure gradient advancing 
the less viscous fluid against the more viscous one, or by gravity, as a 
result of a density difference between the fluids. The interface 
deforms, and different 
modes grow and compete dynamically leading to undulated patterns. 
The physics of the first stage of interface 
instability is captured by 
linear stability analysis~\cite{Rev}.
After this initial linear behavior, 
the system evolves through a ``weakly nonlinear'' stage 
to a complicated late stage, characterized by  
formation of fingers and bubbles, in which nonlinear 
effects dominate. 

The majority of analytical investigations of
the dynamics of fingering instability focus on 
linear stability analysis or else on selection 
of steady-state patterns~\cite{Rev}. 
More recent analytical development~\cite{Tan} addresses the fully nonlinear 
time-evolving flow in the small surface tension limit. 
In contrast, our present work develops an analytical 
approach which links the initial linear behavior with the strongly 
nonlinear advanced stages of the flow. 
We consider early stages of the flow, focusing especially on 
the onset of nonlinear effects. We employ an analytical 
approach known as a mode coupling theory, following 
the method of Haan~\cite{Haa} who studied the Rayleigh-Taylor 
instability in inertial confinement fusion. We previously
applied this method to the Saffman-Taylor instability in radial
flow geometry Hele-Shaw cells~\cite{Mir}.  A closely related study 
in the context of solidification was carried out in 
reference~\cite{Mathur}. Although our approach
is quantitatively accurate only at early stages of pattern formation, we
gain insight into the mechanisms of pattern selection and evolution.

For the rectangular flow geometry, numerical 
simulations~\cite{Try1,Try2,Cas1,Cas2} and 
experiments~\cite{Mah,DiF,Zha} show that 
in the nonlinear regime, the viscosity contrast 
$A$ (defined as the difference 
between the two fluid viscosities divided 
by their sum) plays a major role in the interface evolution. 
Most noteworthy is the role of viscosity contrast in breaking
the symmetry of the interface.
Linear stability analysis does not reveal 
any interface symmetry breaking in the rectangular geometry. In 
contrast, for the radial geometry the distinction between inside and 
outside of the interface always breaks the symmetry.

For the rectangular geometry, in the case of zero 
viscosity contrast ($A=0$), the flow 
is symmetric about the original flat interface position. 
On average, 
the fingers of one fluid penetrate the second fluid as 
much as fingers from the second fluid penetrate the first one. 
For nonzero viscosity 
contrast ($A \neq 0$), the less viscous fluid 
penetrates further into the more viscous one. The less viscous fluid 
fingers exhibit stronger length variation than 
the more viscous fluid fingers. 
For intermediate flow stages, the result is short fingers of 
the more viscous fluid inside the less viscous
fluid, and long fingers of the less viscous 
fluid inside the more. In this 
case, the up-down interface symmetry is obviously broken. At late stages 
the long fingers may pinch off forming bubbles.

In references~\cite{Try1,Try2,Cas1,Cas2} the mathematical description 
of the fingering dynamics is written in terms of nonlinear 
integro-differential equations. These equations cannot be 
solved exactly, and consequently the role of $A$ in asymmetry 
development is primarily revealed by numerically solving these equations. 
A simple analytical description of it remains to be explicitly addressed. 

In addition to the up-down symmetry breaking issue, 
other possible phenomena like finger tip-splitting and 
growth saturation are worth investigating. 
We investigate the saturation of fast-growing modes. By saturation 
we mean a reduction of the exponential rate of growth. 
Numerical studies of Tryggvason and Aref~\cite{Try1,Try2} exhibit a 
saturation for Saffman-Taylor flow. Here, we verify the saturation 
of growth at third order mode coupling, like that 
seen in the Rayleigh-Taylor instability~\cite{Haa,Haa2,Tow}.

Numerical 
simulations for rectangular geometry flow~\cite{DeG,Mei} indicate 
that fingers undergo a type of tip-splitting instability 
in the late stages of interface evolution when the 
dimensionless surface tension parameter $B$~\cite{Cap} 
is sufficiently small. Experiments~\cite{Par2,Tab,Max,Arn}, performed at 
very low $B$, observed that the fingers of the less viscous fluid 
split at their tips while penetrating the more viscous one. 
They also observed highly ramified 
fractal-like structures, which hardly resemble a Saffman-Taylor finger. 
Despite extensive numerical and analytical 
calculations (see, for example, reference~\cite{Tan}), 
considerable uncertainty surrounds the theoretical 
understanding of such splitting events. In contrast, 
for the radial flow geometry, finger tip-splitting is common experimentally, 
and it is predicted by a second-order mode coupling theory~\cite{Mir}. 
In this paper, we demonstrate that symmetries of the rectangular flow 
remove a force driving tip-splitting at second order, explaining 
why it does not commonly occur in early stages of the flow. We investigate 
the possibility of finger tip-splitting at third order, but find that it 
does not occur.

Section 2 carries out our analytical weakly nonlinear investigation
and derives a differential equation describing the early nonlinear
evolution of the interface modes.  In section 3, we interpret the
results obtained in section 2.  We identify and analyze the basic
symmetries of the mode coupling differential equation. The
differential equation exhibits the interface asymmetry discussed
above, and its relation to the value of $A$.  It also indicates the
absence of finger tip-splitting and the existence of
growth saturation at third order. We concentrate our
attention on the coupling of a small number of modes. Section 4
presents our final remarks and perspectives.

\section{The mode coupling differential equation} 
\label{diffeq}

Consider two semi-infinite immiscible viscous fluids, flowing 
in a narrow gap of thickness $b$, in between two parallel 
plates (see figure 1). 
We assume that $b$ is smaller than any other length scale in the problem,
and therefore the system is considered to be effectively two-dimensional. 
Denote the densities and viscosities of the lower 
and upper fluids, respectively as $\rho_{1}$, 
$\eta_{1}$ and $\rho_{2}$, $\eta_{2}$. 
The fluids are assumed to be incompressible, and the flows are assumed to be
irrotational, except at the interface. Between the two fluids there exists
a surface tension $\sigma$. Inject fluid 1 at constant external flow velocity
$\vec v_{\infty}=v_{\infty}\hat{y}$ at $y=-\infty$ and withdraw fluid 2 
at the same velocity at $y=+ \infty$. 
We describe the system in a frame moving 
with velocity $\vec v_{\infty}$, so that the interface may deform, 
but it does not displace from $y=0$ on the average.
In order to include the acceleration of gravity $\vec g$, 
we tilt the cell so that the $y$ axis lies at 
angle $\beta$ from the vertical direction. During the flow, the 
interface has a perturbed shape described as $y = \zeta(x,t)$ over the range 
$0 \le x \le L$ in the comoving frame.

The early nonlinear evolution of the interface obeys a
mode coupling equation. We extend Haan's work on the 
Rayleigh-Taylor problem~\cite{Haa} to the
case of viscous flow in a Hele-Shaw cell. 
The analytic model we seek 
predicts the evolution of the interface perturbation amplitude $\zeta(x,t)$.

We begin by representing the net perturbation $\zeta(x, t)$ in the form 
of a Fourier expansion
\begin{equation}
\label{expansion}
\zeta(x,t)=\sum_{k} \zeta_{k}(t) \exp(ikx),
\end{equation}
where
\begin{equation}
\label{Fourier}
\zeta_{k}(t)=\frac{1}{L} \int\zeta(x,t) \exp(-i k x) dx
\end{equation}
denotes the complex Fourier mode amplitudes.
Expansion~(\ref{expansion}) includes all possible modes $k$, with the 
exception of $k=0$ since we are in a comoving frame. The wave vectors 
are constrained to lie on the $x$ axis, but can be either 
positive or negative. 
We apply periodic boundary conditions in $x$ limiting the wave number $k$ 
to discrete allowed values 
$2 \pi n/L$, for integer $n$. Experimental realization of the Saffman-Taylor 
flow in a cylindrical Hele-Shaw cell~\cite{Zha} showed that periodic 
boundary conditions are similar to the presence of physical sidewalls. 

In the following 
paragraphs we use Fourier analysis to gain insight into the analytic 
structure of the flow dynamics. The Fourier approach is of particular 
interest, since despite its simplicity, it allows one to adopt the 
familiar physical ideas of modes and growth rates developed in the 
linear theory. Here we go beyond the level of linear 
stability analysis and analytically investigate the weakly nonlinear 
stage of the evolution.

Since we are interested in the early nonlinear behavior of the system,
our first task is to derive a differential equation for $\zeta_{k}$, correct 
to third order.
The relevant hydrodynamic equation is Darcy's law~\cite{Saf,Rev}
\begin{equation}
\label{Darcy}
\eta_{i}(\vec v_{i} + \vec v_{\infty})= -\frac{b^2}{12} \left \{ \vec\nabla p_{i} - \rho_{i} (\vec g \cdot \hat{y})\hat{y}\right \},
\end{equation}
where $\vec v_{i}=\vec v_{i}(x,y)$ 
and $p_{i}=p_{i}(x,y)$ are, 
respectively, the velocity and pressure in fluids 
$i=1$ and $2$. 
Equation~(\ref{Darcy}) 
derives from the Navier-Stokes equation by neglecting 
inertial terms, assuming a parabolic flow
profile with vanishing velocity at the plates, and by averaging the velocity
over the perpendicular direction to the $x-y$ plane.

Supplement equation~(\ref{Darcy}) 
with the irrotational flow condition $\vec \nabla \times \vec v_{i}=0$.
Under this circumstance the velocity is given by $\vec v_{i}=-\vec \nabla 
\phi_{i}$, where $\phi_{i}$ represents the velocity 
potential in each of the fluids. Rewrite equation~(\ref{Darcy}) in terms 
of velocity potentials and then integrate. After dropping an arbitrary 
constant of integration we write Darcy's law for velocity potential
\begin{equation}
\label{Darcypot}
\eta_{i}\phi_{i}= \frac{b^2}{12} \left \{ p_{i} + \rho_{i} g y \cos \beta \right \} + \eta_i v_{\infty} y.
\end{equation}

Rewrite equation~(\ref{Darcypot}) for each of the fluids and 
subtract the resulting equations from each other to obtain the 
jump condition
\begin{equation}
\label{combined}
A \left ( \phi_{1}|_{y=\zeta} + \phi_{2}|_{y=\zeta} \right ) -  \left ( \phi_{1}|_{y=\zeta} - \phi_{2}|_{y=\zeta} \right ) = 2 \left \{ \frac{b^2(p_{2} - p_{1})}{12(\eta_{1} + \eta_{2})} + Uy \right \}
\end{equation}
where    
\begin{equation}
\label{contrast}
A=\frac{\eta_{2} - \eta_{1}}{\eta_{2} + \eta_{1}}
\end{equation}
is the viscosity contrast and 
\begin{equation}
\label{U}
U=\frac{b^{2}(\rho_{2} - \rho_{1}) g \cos \beta}{12(\eta_{1} + \eta_{2})} + Av_{\infty}
\end{equation}
is a characteristic velocity of the problem. 
We concentrate attention on $U>0$, because there is no instability when $U<0$.

The pressure difference at the interface between the two fluids 
is given by~\cite{Rev}
\begin{equation}
\label{pressure}
(p_{2} - p_{1})|_{y=\zeta} = \sigma \left ( \frac{1}{R_{\|}} + \frac{1}{R_{\bot}} \right ).
\end{equation}
The two principal radii of curvature, $R_{\|}$ and $R_{\bot}$, describe the interface locally. The curvature in the $x-y$ plane is~\cite{Dub}
\begin{equation}
\label{parallel}
\kappa_{\|}=\frac{1}{R_{\|}}=\left ( \frac{\partial^2 \zeta}{\partial x^{2}} \right ) \left [ 1 + \left(\frac{\partial\zeta}{\partial x} \right)^2 \right ]^{-\frac{3}{2}}.
\end{equation}
Meanwhile $R_{\bot} \sim b/2$, the radius of curvature in the direction 
perpendicular to the parallel plates, is controlled by the 
contact angle of the two-fluid interface at the plates. 
Typically, one finds that $R_{\bot}$ is nearly constant~\cite{McL,Par}. 
Even though $1/R_{\bot} \gg 1/R_{\|}$, the perpendicular curvature does 
not significantly affect the motion in our problem, since its 
gradient is nearly zero.

Combine Darcy's law equation~(\ref{combined}), 
with equations~(\ref{pressure}) 
and~(\ref{parallel}) for the pressure difference and rescale
lengths by $L$ and time by $L/U$, to obtain the dimensionless 
equation of motion
\begin{equation}
\label{dimensionless2}
A \left ( \phi_{1}|_{y=\zeta} + \phi_{2}|_{y=\zeta} \right ) -  \left ( \phi_{1}|_{y=\zeta} - \phi_{2}|_{y=\zeta} \right ) = 2 \left [\zeta + B\kappa_{\|} \right ].
\end{equation}
where
\begin{equation}
\label{B}
B=\frac{b^{2} \sigma}{12 U (\eta_{1} + \eta_{2}) L^{2}}
\end{equation}
is a dimensionless surface tension coefficient. 
From now on 
we work, unless otherwise stated, with the dimensionless 
equation of motion.

For our weakly nonlinear analysis we are 
interested in third order contributions in the perturbation amplitudes.
Therefore, all the quantities in equation~(\ref{dimensionless2}) are 
evaluated at the perturbed interface $y=\zeta(x,t)$, and
not at the initial interface position $y=0$ as is usually done in 
linearized surface deformation problems. The 
nonlinear terms arise because of this important distinction.

The incompressibility of the fluids $(\vec \nabla \cdot \vec v_{i}=0)$ 
means the velocity potentials satisfy Laplace's equation 
$\nabla^{2}\phi_{i}=0$. For points far from the interface, we expect 
uniform, steady flow uninfluenced 
by the interface. Therefore, we require
that the system evolves with uniform 
velocity $\vec v_{\infty}=v_{\infty}\hat{y}$ 
in the limits $y \rightarrow \pm \infty$. 
Recall that $\vec v_{i}$ are measured in the comoving frame, and thus 
vanish at $y \rightarrow \pm \infty$. 
So, $\phi_{1}$ and $\phi_{2}$ go to constants (that we set to zero) as y goes 
to $-\infty$ and $+\infty$, respectively.

Now define Fourier expansions for the velocity
potentials $\phi_{i}$, which must obey Laplace's equation 
 $\nabla^{2}\phi_{i}=0$, the boundary conditions 
at $y \rightarrow \pm \infty$, and periodic boundary conditions on $x$. 
The general velocity potentials obeying
these requirements are
\begin{equation}
\label{phi1}
\phi_{1}=\sum_{k \neq 0} \phi_{1 k}(t) \exp(|k|y) \exp(ikx),
\end{equation}
and
\begin{equation}
\label{phi2}
\phi_{2}=\sum_{k \neq 0} \phi_{2 k}(t) \exp(-|k|y) \exp(ikx).
\end{equation}
Substitute expansions~(\ref{phi1}) and ~(\ref{phi2}) into 
the equation of motion~(\ref{dimensionless2}). Keep third order terms 
in the perturbation amplitudes, and then Fourier transform. 
For example, the Fourier transform of the lower fluid velocity potential at 
the perturbed interface $y=\zeta$ with wave vector $k$ takes the form
\begin{equation}
\label{Darcy3}
\hat{\phi}_{1}(k)=\phi_{1k}(t) + \sum_{k' \neq 0} |k'|\phi_{1k'}(t) \zeta_{k - k'} + \frac{1}{2} \sum_{k',q \neq 0} (k')^{2} \phi_{1k'}(t) \zeta_{q}\zeta_{k - k' - q},
\end{equation}
where $\zeta_{k}$ is the Fourier expansion of $\zeta$, given by 
equation~(\ref{Fourier}). A similar expression for $\phi_{2}|_{y=\zeta}$ can 
be easily obtained. 
Likewise, the Fourier transform of the in-plane curvature with wave vector $k$, 
valid up to third order in $\zeta$, is
\begin{equation}
\label{newequation}
\hat{\kappa}_{\|}(k)= -k^{2}\zeta_{k} - \frac{3}{2} \sum_{k',q \neq 0} (k')^{2} q [k - k' - q] \zeta_{k'}\zeta_{q}\zeta_{k - k' - q}.
\end{equation}

To close equation~(\ref{dimensionless2}) we need 
additional relations expressing 
the velocity potentials in terms of the perturbation amplitudes. 
To find these, consider 
the kinematic boundary condition, which states that
the normal components of each fluid's velocity at the interface equals 
the normal velocity of the interface itself~\cite{Ros}, i.e.
\begin{equation}
\label{b.c.}
\frac{\partial \zeta}{\partial t}= \left ( \frac{\partial \zeta}{\partial x}\frac{\partial \phi_{i}}{\partial x} \right )_{y=\zeta} - \left (\frac{\partial \phi_{i}}{\partial y} \right )_{y=\zeta}.
\end{equation}
Expand equation~(\ref{b.c.}) to third order in $\zeta$ and then
Fourier transform. Solving for $\phi_{ik}(t)$ consistently to third order 
in $\zeta$ yields
\begin{eqnarray}
\label{phi1t}
\phi_{1k}(t)& = & -\frac{\dot{\zeta}_{k}}{|k|} + \sum_{k' \neq 0} sgn(kk')\dot{\zeta}_{k'}\zeta_{k - k'}\nonumber \\
            & - & \sum_{k',q \neq 0} \frac{kq}{|k|}sgn(k'q)\dot{\zeta}_{k'}\zeta_{q - k'}\zeta_{k - q} + \sum_{k',q \neq 0}\frac{k'}{|k|} \left ( k - q - \frac{k'}{2} \right )\dot{\zeta}_{k'}\zeta_{q}\zeta_{k - k' - q}\nonumber \\
\end{eqnarray}
and a similar expression for $\phi_{2k}(t)$. 
Here $sgn$ denotes the sign function. For instance, $sgn(kk')=1$ if 
$(kk')>0$ and $sgn(kk')=-1$ if $(kk')<0$. The overdot denotes total 
time derivative.

Substitute this last expression for $\phi_{1k}(t)$ into 
equation~(\ref{Darcy3}), and again keep only 
cubic terms in the perturbation amplitude (the same procedure must 
be repeated for fluid 2). For $k \neq 0$, 
Darcy's law~(\ref{dimensionless2}) becomes the differential equation
\begin{eqnarray}
\label{result}
\dot{\zeta}_{k}& = & \lambda(k)\zeta_{k} + A |k| \sum_{k' \neq 0} \left[ 1 - sgn(kk') \right] \dot{\zeta}_{k'}\zeta_{k - k'}\nonumber \\
               & + & \sum_{k',q \neq 0} |k| |q| sgn(k'q) \left[ 1 - sgn(kq) \right] \dot{\zeta}_{k'}\zeta_{q - k'}\zeta_{k - q}\nonumber \\
               & + & \sum_{k',q \neq 0}k' \left[ k - q - \frac{k'}{2} - \frac{|k'||k|}{2k'} \right ] \dot{\zeta}_{k'}\zeta_{q}\zeta_{k - k' - q}\nonumber \\
               & - & \frac{3}{2} B \sum_{k',q \neq 0} |k| (k')^{2} q [k - k' - q] \zeta_{k'}\zeta_{q}\zeta_{k - k' - q}, \nonumber \\
\end{eqnarray}
where $A$ is the viscosity contrast as defined in equation~(\ref{contrast}),
$B$ is the dimensionless surface tension coefficient given in 
equation~(\ref{B}) and
\begin{equation}
\label{growth}
\lambda(k)=|k|(1 - Bk^{2})
\end{equation}
is the linear growth rate.

Equation~(\ref{result}) is the mode coupling equation of the Saffman-Taylor 
problem for the rectangular geometry Hele-Shaw flow. It gives us the time 
evolution of the perturbation amplitudes $\zeta_{k}$ accurate to third order, 
conveniently written in terms of the two dimensionless parameters 
$A$ and $B$. The first term on the 
right-hand side of equation~(\ref{result}) reproduces 
the linear stability analysis [1-3]. 
The second term, of great importance for understanding 
the interface asymmetry, represents second-order mode coupling. The remaining 
terms depict the third order contributions to the mode coupling equation. 
They lead to saturation of the growth compared with the linear equation 
of motion. Beyond third order, we anticipate that all even terms are 
multiplied by $A$, and $B$ enters only into odd terms. 
In the following section we investigate the mode coupling 
equation~(\ref{result}) in more detail.

\section{Discussion}

We begin our discussion by analyzing the basic symmetries present in
the Saffman-Taylor problem and considering how they constrain the form
of the mode coupling equation~(\ref{result}). Then we look in detail
at features of the interface morphology dictated by the first-,
second- and third-order terms.

Several symmetry operations leave the Hele-Shaw cell invariant. Because
of the periodic boundary conditions introduced in
section~\ref{diffeq}, the cell is invariant under infinitesimal
horizontal translations $T_x^{\epsilon}: (x,y) \rightarrow (x+\epsilon,y)$.
Provided the cell is sufficiently long in the vertical direction, we
may assume vertical translational symmetry $T_y^{\epsilon}: (x,y)
\rightarrow (x,y+\epsilon)$. Additional symmetries of the cell are the
vertical mirror $M_v: (x,y) \rightarrow (-x,y)$ and the horizontal
mirror $M_h: (x,y) \rightarrow (x,-y)$.

However, for the Saffman-Taylor problem, the differing fluid densities
and viscosities, the gravitational acceleration $\vec g$ and external
flow $\vec v_{\infty}$ break the symmetry $M_h$. We define a new
symmetry transformation of the Saffman-Taylor problem $\tilde{M}_h:
(x,y,A,B) \rightarrow (x,-y,-A,B)$. Interchanging viscosities $\eta_1$
and $\eta_2$ reverses the sign of $A$ (see equation~(\ref{contrast})).
To understand the transformation of $B$ defined in equation~(\ref{B}) we
must study the transformation of $U$ defined in equation~(\ref{U}).
Interchanging the densities $\rho_1$ and $\rho_2$, and simultaneously
reversing the direction of gravitational acceleration $\vec g$ leaves the
first term of equation~(\ref{U}) for $U$ invariant. Reversing the sign
of $A$ and simultaneously reversing the direction of external flow 
velocity $\vec v_{\infty}$ leaves invariant the second term 
of equation~(\ref{U}).
Invariance of $U$ implies invariance of $B$. Thus $\tilde{M}_h$ swaps
the two fluids and reverses the direction of gravitational
acceleration and external flow at the same time as it reflects the $y$
coordinate.

The Fourier modes {$\zeta_k \exp(ikx)$} are basis functions for
representations of the symmetry group generated by the above
operations. By investigating the transformation of modes under group
generators we can check the symmetry properties of the mode coupling
equation~(\ref{result}), and explain the presence and/or absence of
various terms. For example, applying horizontal translation
$T_x^{\epsilon}$ to a mode of wave vector $k$ multiplies its
coefficient $\zeta_k$ by the complex phase factor $\exp(ik\epsilon)$.
All terms in the mode coupling equation must transform
identically under this translation. Every product of coefficients
$\prod\limits_j \zeta_{k_j}$ gets multiplied by 
$\exp(i \sum\limits_{j} k_{j} \epsilon)$.
Consequently, the wave vectors {$k_j$} in each term must sum up
to a common value, $k$. 
Inspection of equation~(\ref{result}) verifies this rule.

Applying vertical translation $T_y^{\epsilon}$ has no effect (the mode
transforms as the identity) because $\zeta(x)$ is defined as the
interfacial height {\em relative} to the unperturbed interface. Thus
the vertical translation symmetry does not significantly constrain the form of
equation~(\ref{result}).

The vertical mirror $M_v$ reverses the sign of $k$. Invariance under
this operation requires that $\zeta_{-k}$ obeys the same equation as
$\zeta_k$. Inspection of equation~(\ref{result}) confirms that all
coefficients are even under the reversal of sign of all wave vectors.
For example, expressions that are first or third order in the
wave vectors are written as an even power of the wave vectors times an
absolute value of a wave vector. The even symmetry of the coefficient
$\lambda(k)$ is explicitly revealed in equation~(\ref{growth}).

The horizontal mirrors (both $M_h$ and $\tilde{M}_h$) reverse the sign
of $\zeta_k$. We will discuss the transformation of individual terms
in equation~(\ref{result}) under these mirror operations in subsequent
sections as we discuss evolution of the interface up to first-,
second-, and third-order in our mode coupling theory.

\subsection{First order - Linear evolution}
\label{first_order}

The linear stability analysis of the viscous fingering problem has
been studied since the late 50's~\cite{Saf,Rev}.  Due to its
importance, we briefly review some features of the linear regime.
Start with the first order solution to equation~(\ref{result})
\begin{equation}
\label{linear}
\zeta_{k}^{lin}(t)=\zeta_{k}(0)\exp[\lambda(k)t].
\end{equation}
The linear growth rate $\lambda(k)$ (see equation~(\ref{growth})),
which leads to exponential growth at small amplitudes, is plotted in
figure 2 for $B=1$. From figure 2 we see that, for small wave vectors,
perturbations grow in time, deforming the interface. At large wave
vectors, surface tension stabilizes short wavelength deviations. From
the linear growth rate~(\ref{growth}) we can extract two important
parameters: the critical wave vector (defined by setting
$\lambda(k)=0$)
\begin{equation}
\label{critical}
k_{c}=\frac{1}{\sqrt{B}}
\end{equation}
beyond which all modes are linearly stable; and the fastest growing
mode (defined by setting $d\lambda(k)/dk=0$)
\begin{equation}
\label{fastest}
k^{*}=\frac{1}{\sqrt{3B}},
\end{equation}
which dominates the initial dynamics of the interface.  The fastest
growing wavelength $\lambda^{*}=2 \pi/k^{*}$ sets a characteristic
length to the problem, giving the experimentally observed period of
the fingering pattern at initial stages of the flow.

Consider the transformation of the first-order terms in our
mode-coupling equation~(\ref{result}) under horizontal reflections.
The coefficient $\zeta_k$ reverses sign on both sides of the equation.
What about $\lambda(k)$? Since it depends only on constant parameters
defining the system under study, independent of the shape of the
interface, it exhibits invariance under $M_h$. As a result the linear
analysis neither predicts, nor explains, the interface asymmetry
observed in simulations~\cite{Try1,Try2,Cas1,Cas2} and
experiments~\cite{Mah,DiF,Zha}. Although at first-order the interface
shapes remain (statistically) symmetric under $M_h$, the
Saffman-Taylor problem does not respect $M_h$ symmetry because of the
distinction between the lower and upper fluids, and the directions of
gravitational acceleration and external flow. The symmetry operation
$\tilde{M}_h$ is the appropriate symmetry for the Saffman-Taylor
problem. Since $B$ is invariant under $\tilde{M}_h$, so is the
linear growth rate $\lambda(k)$, confirming the required symmetry.

It is convenient to rewrite the net perturbation~(\ref{expansion}) in
terms of cosine and sine modes
\begin{equation}
\label{sincos}
\zeta(x,t)= \sum_{k > 0}  \left[ a_{k}(t)\cos(kx) + b_{k}(t)\sin(kx) \right ],
\end{equation}
where $a_{k}=\zeta_{k} + \zeta_{-k}$ and $b_{k}=i \left ( \zeta_{k} -
\zeta_{-k} \right )$ are real-valued. Tryggvason and Aref~\cite{Try2}
studied the nonlinear behavior of the interface numerically,
considering the interaction between two cosine waves. Similarly, to
illustrate the linear evolution of the interface, we perturb the
initially flat interface with two cosine modes $a_{k_s}$ and
$a_{k_f}$. We consider a dominant {\em fundamental} wave of wave
vector $k_f=k^{\ast}=1/\sqrt{3}$ and initial amplitude $a_{k_f}(0)$,
and a second {\em sub-harmonic} wave of wave vector $k_s=k_{f}/2$ and
relatively weak amplitude $a_{k_s}(0)=0.2 a_{k_f}(0)$.

Exactly two wavelengths of the fundamental mode fit in the Hele-Shaw
cell, so the fundamental mode is invariant under horizontal
translations $T_x^{L/2}$. The sub-harmonic perturbation $a_{k_s}$
breaks this translational symmetry of the fundamental perturbation
$a_{k_f}$ by slightly altering the relative lengths of the two
upward-pointing fingers. Up-down symmetry of the interface is weakly
broken, by construction, due to our arbitrary choice of sub-harmonic
$a_{k_s}\cos{(k_s x)}$. Had we taken the sub-harmonic
$b_{k_s}\sin{(k_s x)}$ instead, the length alternation would have
appeared in the downward-pointing fingers.

Figure 3a depicts the interface given by linear theory. We observe
that the interface remains nearly up-down symmetric, showing no
particular tendency to interface asymmetry development. Each uncoupled 
mode by itself respects up-down symmetry. The
translational and reflection symmetry breaking of the interface are
artifacts of the particular initial condition we select.
The interfacial
deformation grows sufficiently large that
quantitative accuracy of any perturbative approach is doubtful. We illustrate
such large amplitudes deliberately, however, because they enhance the
visibility of nonlinear effects described in the next two subsections.

\subsection{Second order - Role of the asymmetry parameter $A$}
\label{second_order}

For sufficiently short times we presume that the perturbation series
defining the mode coupling equation of motion, and its solutions,
converge to their true forms as successively higher-order terms are
incorporated. Truncating equation~(\ref{result}) at second order
should then result in quantitatively small changes in the calculated
interface profile $\zeta(x,t)$. Although quantitatively small, these
changes should incorporate the principal corrections to the linear
interface evolution discussed above in section~\ref{first_order}.
Indeed, this section explains two important qualitative features of
the interfacial evolution: up-down symmetry breaking, and the general 
absence of finger splitting.

We consider first the breaking of up-down symmetry. As noted above, in
section~\ref{first_order}, the linear evolution respects reflection
symmetry $M_h$ since $\zeta_k$ reverses sign on both sides of the
equation of motion, while $\lambda(k)$ is unaffected. In contrast, the
second order term in equation~(\ref{result}) breaks $M_h$ symmetry
because it does not change sign. The second order term is multiplied
by the viscosity contrast $A$. Hence the breaking of up-down symmetry
depends upon the difference between the viscosities of the two fluids. 
In equation~(\ref{result}) the role of $A$ in asymmetry 
development is clearly identified, being revealed without solving 
any complicated nonlinear integro-differential 
equations~\cite{Try1,Try2,Cas1,Cas2}. Our mode coupling approach provides 
a transparent and simple way of identifying the intrinsically 
nonlinear character of the viscosity contrast $A$.

Symmetry breaking in the equation of motion translates into symmetry
breaking in the solution. We now solve equation~(\ref{result}) to
second order accuracy. Substitute the linear solution~(\ref{linear})
into the second-order terms on the right hand side of
equation~(\ref{result}), to obtain
\pagebreak
\begin{eqnarray}
\label{try}
\dot{\zeta}_{k}& = & \lambda(k)\zeta_{k} \nonumber \\
               & + & |k| A \sum_{k' \neq 0} \left[ 1 - sgn(kk') \right] \lambda(k')\zeta_{k'}(0)\zeta_{k - k'}(0) \exp[(\lambda(k - k') + \lambda(k'))t]
\nonumber \\ 
               & + & {\cal O} (\zeta_{k}^{3}),
\end{eqnarray}
where $\lambda(k - k')$ and $\lambda(k')$ are the linear growth rates
related to the modes $k - k'$ and $k'$, respectively.
Equation~(\ref{try}) is a standard first order differential
equation~\cite{Gra} with the solution
\begin{equation}
\label{try2}
\zeta_{k}=\zeta_{k}^{lin} + |k| A \sum_{k' \neq 0} \zeta_{k'}(0)\zeta_{k - k'}(0) {\cal G}(k,k',t) +  {\cal O} (\zeta_{k}^{3}),
\end{equation}
where
\begin{eqnarray}
\label{F}
{\cal G}(k,k',t)& = &\left[ 1 - sgn(kk') \right] \left ( \frac{\lambda(k')}{\lambda(k - k') + \lambda(k') - \lambda(k)} \right ) \nonumber \\
         & \times & \left( \exp[\lambda(k - k') + \lambda(k')]t - \exp[\lambda(k)t] \right).
\end{eqnarray}
The apparent singularities in ${\cal G}(k,k',t)$ at
$\lambda(k)=\lambda(k') + \lambda(k - k')$ are cancelled by zeros in
the numerator of~(\ref{F}), and each term in the sum
(equation~(\ref{try2})) is regular at these points. At this apparent
``resonance point'' ${\cal G}(k,k',t)$ varies as $t\exp[\lambda(k)t]$.

We use the second order solution~(\ref{try2}) to investigate the
coupling of a small number of modes.  As discussed in
reference~\cite{Try2}, even the interaction of two modes can lead to
patterns relevant to more complex statistical-fingering
calculations~\cite{Try1}. Our discussion will be simpler if we replace
the complex Fourier modes $\zeta_k$ with sine and cosine modes as in
equation~(\ref{sincos}). The second order equations of motion are
\begin{equation}
\label{cosine}
\dot{a}_{k}=\lambda(k)a_{k} + A k \sum_{k'>0} \left[ \dot{a}_{k'}a_{k + k'} + \dot{b}_{k'}b_{k + k'} \right ],
\end{equation}
\begin{equation}
\label{sine}
\dot{b}_{k}=\lambda(k)b_{k} + A k \sum_{k'>0} \left[ \dot{a}_{k'}b_{k + k'} - \dot{b}_{k'}a_{k + k'} \right ].
\end{equation}
Note how the products of sine and cosine amplitudes are arranged to
preserve the $M_v$ symmetry under which the sine functions are odd and
the cosines are even. Solutions to these equations are similar in
form to~(\ref{try2}).

Figure 3b illustrates the second-order solution taking the same
two-mode initial conditions used in figure 3a. Since $k_s=k_f/2$,
equations~(\ref{cosine}) and~(\ref{sine}) couple the growth of the
sub-harmonic to the amplitude of the fundamental. Finger competition,
associated with the sub-harmonic mode, is enhanced to a degree
proportional to the viscosity contrast $A$. Figure 3b represents the
extreme case $A=1$. Length variation between the upwards fingers is
stronger than in figure 3a.  Notice also that the central upward
finger advances more strongly into the upper fluid than the downward
fingers into the lower fluid. The upward finger at $x=0$ advances less 
strongly.

Mode coupling not only influences the {\em magnitude} of the
sub-harmonic, but also selects its {\em phase}. Without loss of
generality we may take $a_{k_f} > 0$ and $b_{k_f}=0$, as we do in the
initial conditions for figure 3b. Now compare the growth rates
of sine and cosine sub-harmonic modes
\begin{equation}
\label{illustration1}
\dot{a}_{k_s}=\lambda(k_s){a}_{k_s} + A k_{s}\dot{a}_{k_s}a_{k_f},
\end{equation}
\begin{equation}
\label{illustration2}
\dot{b}_{k_s}=\lambda(k_s){b}_{k_s} - A k_{s}\dot{b}_{k_s}a_{k_f}.
\end{equation}
Positive viscosity contrast $A>0$ increases the growth rate of the
cosine sub-harmonic $a_{k_s}$, causing increased variability among the
lengths of fingers of the less viscous lower fluid 1 penetrating
downwards into the more viscous upper fluid 2. Note that the sign of
$a_{k_s}$ is dictated by initial conditions, and not influenced by
mode-coupling.  Reversing the sign of $a_{k_s}$ has the effect of
interchanging which of the two upwards pointing fingers will grow at
the expense of the other. This is tantamount to a horizontal
translation by $L/2$.  Finally consider patterns with $a_{k_s}=0$ and
$b_{k_s} \ne 0$.  Inspecting equation~(\ref{illustration2}) we see
that $A>0$ inhibits the growth of sine modes $b_{k_s}$. Sine modes
would vary the lengths of fingers of the more viscous upper fluid 2
penetrating into the less viscous lower fluid 1, but their growth is
inhibited.

Reversing the sign of $A$ exactly reverses the above conclusions.
Sub-harmonic sine modes will be favored over cosine modes. This happens
because of the $\tilde{M}_h$ invariance of the Saffman-Taylor problem.
Reversing the signs of both $\zeta(x)$ and $A$ leaves the form of
equation~(\ref{result}) invariant. Consequently, a randomly chosen
up-down symmetric (on average) initial condition always evolves into a
symmetry-broken interfacial pattern in which the fingers of the less
viscous fluid exhibit variable finger lengths penetrating into the
more viscous fluid~\cite{Saf,Mah,DiF,Zha}.

Further inspection of the cosine and sine mode coupling
equations~(\ref{cosine}) and~(\ref{sine}) reveals that, while the
presence of large wave number modes influences the growth of smaller
wave number modes, the reverse is not true. For example, the growth of
any mode $k$ cannot be influenced (up to second order) by modes of
only smaller wave numbers.  In particular, there is no second order
term entering the equation of motion for the fundamental mode ${k_f}$
in the presence of the sub-harmonic $k_s$. Likewise the presence of
$k_f=k^{*}$ and $k_s=k_f/2$ cannot alter the evolution of the {\em
harmonic} mode $k_h=2k_f$. Since $k_h > k_c$, the harmonic mode is
linearly stable $(\lambda(k_h)<0)$ and will not spontaneously grow.

This observation yields insight into the absence of finger
tip-splitting in the rectangular geometry Saffman-Taylor problem, at
second order.  Splitting the tips of the fingers in figure 3b would
require the presence of a sizable harmonic mode of wave number $k_h$.
Even if such modes enter weakly through initial conditions or random
noise, they quickly die out.

This is in striking contrast to the radial geometry Saffman-Taylor
problem~\cite{Mir}, where mode coupling drives the growth of harmonic
modes. In the radial case we may talk about the fundamental cosine
mode number $n_f$, where $n_f$ is an integer counting the number of
oscillations around the growing perimeter. The influence of a
fundamental mode $n_f$ on the growth of its harmonic mode $n_h=2n_f$
is given by the equation of motion~\cite{Mir}
\begin{equation}
\label{radial}
\dot{a}_{n_h}=\lambda(n_h)a_{n_h} + {\cal C} a_{n_f}^2,
\end{equation}
where $\lambda(n_h)$ denotes the linear growth rate related to the
mode $(n_h)$ in the radial geometry case, and ${\cal C}$ is
negative.  Even if $\lambda(n_h) < 0$, the harmonic mode can still
grow provided the fundamental mode is present, $a_{n_f} \ne 0$. In the
radial geometry the presence of the fundamental forces growth of the
harmonic, while in the rectangular geometry the fundamental does not
influence the harmonic at second order. Because ${\cal C}<0$,
$a_{n_h}$ is driven negative, the sign that is required to cause
fingers to split.

Consider the behavior of the coefficient of $a_{n_f}^2$ under the
radial analogue of $\tilde{M}_h$ (here denoted by
$\tilde{M}_h^{radial}$), which interchanges the two fluids and
reverses the direction of external flow while reversing the sign of
the interfacial perturbation $\zeta$. The coefficient, ${\cal C}$,
contains both even and odd terms under $\tilde{M}_h^{radial}$. However,
the odd terms are vanishingly small compared to the even terms for
large unperturbed radii of curvature. We will treat ${\cal C}$ as
effectively even under $\tilde{M}_h^{radial}$ in the discussion.

To transform consistently under $\tilde{M}_h^{radial}$, the radial
equation of motion~(\ref{radial}) must contain only terms that are
odd. The second order term, however, is even and thus breaks
$\tilde{M}_h^{radial}$ symmetry. This is expected, because
$\tilde{M}_h^{radial}$ is not a symmetry of the radial flow
Saffman-Taylor problem. The interface is always in-out asymmetric in
radial flow because we can always distinguish the region of space that
lies inside the interface from the region that lies outside. For
instance, the outward radial motion in which air pushes oil in a
radial Hele-Shaw cell (divergent flow) is not equivalent to the inward
radial motion corresponding to withdrawal of oil surrounded by air
(convergent flow)~\cite{Tho}. This broken symmetry allows finger
tip-splitting.

In contrast, $\tilde{M}_h$ {\em is} a symmetry of the {\em
rectangular} Saffman-Taylor problem. Thus, even terms in the equation
of motion, like the one that causes tip-splitting in the radial
geometry, are forbidden in the rectangular geometry. Indeed, it can be
shown that the amplitudes of these terms vanish when taking the
rectangular geometry (large radius) limit of the radial flow equations
of motion~\cite{Mir}. At second order the driving force creating
finger tip-splitting is eliminated by the special symmetry of the
rectangular geometry.

While $\tilde{M}_h$ rules out terms like ${\cal C} a_{k_f}^2$ with
${\cal C}$ even under $\tilde{M}_h$, it does not rule out similar
terms with odd coefficients such as the viscosity contrast $A$. From
our mode coupling equations~(\ref{illustration1})
and~(\ref{illustration2}) we know they are absent, because small wave
number modes do not influence the growth of large wave number modes.
We lack a symmetry explanation for why this is so.

Intermediate cases, between radial and rectangular flow geometry, are
provided by the wedge geometry~\cite{Tho,Ama,Tu}.  In the wedge
geometry the fluids flow in a Hele-Shaw cell in which the side walls
form a wedge with an opening angle of $\theta_{0}$ $(-2 \pi \leq
\theta_{0} \leq 2 \pi)$, where  $\theta_{0} > 0$ ($\theta_{0} < 0$)
corresponds to a divergent (convergent) flow.
Experiments in the wedge geometry~\cite{Tho} with $A > 0$
observed an increasing sensitivity to finger tip-splitting for 
larger angle $\theta_{0} > 0$. A related experiment~\cite{Tho} 
with $A < 0$ and convergent flow shows that fingers grow, but 
tip-splitting of inward fingers is inhibited. At second order, the specially 
symmetric case $\theta_0 = 0$ 
is unique in the absence of a driving force leading to tip splitting.

\subsection{Third order - Onset of saturation and absence of tip-splitting}
\label{third_order}

Now we examine the whole third order mode coupling
equation~(\ref{result}), taking into account the contributions coming
from the third order terms. Since the final expressions are somewhat
complicated, we start our discussion by considering the evolution of a
single mode. The third order mode coupling equation~(\ref{result})
reduces to
\begin{equation}
\label{single}
\dot{\zeta}_{k}=\lambda(k)\zeta_{k} + 
k^{3} \left [ \frac{5}{2} B k^{2} - 1 \right ] \zeta_{k}^{3},
\end{equation}
where we have replaced $\dot{\zeta}_{k}$ with $\lambda(k)\zeta_{k}$ in
terms already of third order on the right hand side of
equation~(\ref{result}). Since we are interested in the fastest
growing mode $k=k^{*}=1/\sqrt{3 B}$, we see from~(\ref{single}) that
the third order terms lead to a saturation of the growth because the
coefficient of $\zeta_{k}^{3}$ is negative. The exponential growth of
the linear instability does not proceed unchecked.

Figure 3c illustrates the full solution up to third order of
equation~(\ref{result}), taking the same initial conditions as were
used in figures 3a and 3b, and including modes $a_{k_f/2}$ and $a_{k_f}$. 
The main effect that is apparent, in comparison with figure 3b, is
significantly diminished amplitude of the fundamental mode $a_{k_f}$
caused by the saturation effect described by equation~(\ref{single}).
There is a slightly increased amplitude of the sub-harmonic $a_{k_s}$.
We also note a slight broadening of the dominant central finger, and
narrowing of the smaller finger at $x=0$.

It is interesting to investigate the possibility of finger
tip-splitting at third order. Finger tip-splitting is associated to the 
magnitude of the harmonic mode $2k_{f}$. It turns out that, at third 
order the cosines modes $a_{k_{f}}$ and $a_{k_{f}/2}$ force growth 
of modes $a_{2k_{f}}$ and $a_{3k_{f}/2}$. 
We consider initial conditions similar to those used in figures 3a-c,
assuming that modes 
$k_f$, $k_s=k_f/2$ are initially present. We study how these
two initial modes force growth of modes 
$2k_f$ and $3k_f/2$. The influence of the fundamental and sub-harmonic 
on the growth of the first harmonic $k_h = 2k_f$ may be expressed by
\begin{equation}
\label{third3}
\dot{a}_{k_h}=\lambda_{eff} a_{k_h}
- \frac{3}{8}Bk_{h}k_{s}^2k_{f} \left [ k_{f} + 2k_{s} \right ]a_{k_f}a_{k_s}^{2} + {\cal O}(a_{k_h}^3),
\end{equation}
where
\begin{equation}
\label{lambda_eff}
\lambda_{eff} = \lambda(k_h) + \frac{k_{f}^{2}k_{h}}{2} \left [ B \left (k_{f}^{2} + \frac{3}{2}k_{h}^{2} \right ) - 1 \right ] a_{k_f}^2 + \frac{k_{s}^{2}k_{h}}{2} \left [ B \left (k_{s}^{2} + \frac{3}{2}k_{h}^{2} \right ) - 1 \right ]a_{k_s}^2.
\end{equation}
In equation~(\ref{third3}), the linear growth rate of mode $k_h$
is increased by the presence of $k_f$ and $k_s$. Although $\lambda(k_h)$
is negative, opposing the growth of the harmonic, the fundamental and
sub-harmonic make $\lambda_{eff}$ less negative. Quantitatively, the role 
of the fundamental dominates the sub-harmonic in increasing $\lambda_{eff}$.

It appears at a glance in equation~(\ref{lambda_eff}) that sufficiently 
large $a_{k_f}$ drives $\lambda_{eff}$ positive, permitting growth of the 
harmonic ${a}_{k_h}$. It turns out that is an artifact of truncating 
the mode coupling theory. If higher order terms in $a_{k_f}$ are kept, 
$\lambda_{eff}$ will remain negative. Terms involving $a_{k_f}$ and $a_{k_s}$ 
make $\lambda_{eff}$ less negative but cannot make it go positive. 
The physical reason that prevents 
positive values for $\lambda_{eff}$ can be understood by considering 
the contour length of the interface. Introducing the harmonic always 
increases the contour length, although the larger the amplitude 
of the fundamental, the smaller the increase upon introducing the harmonic. 
Multiplying the contour length by the surface tension yields surface 
energy that favors minimum contour length. 

There will be a small amplitude 
of the harmonic present due to the driving term proportional to 
$a_{k_f}a_{k_s}^{2}$ in equation~(\ref{third3}), but it is doubtful 
the amplitude will be sufficiently large to split the fingers, because 
it varies like the second power of the sub-harmonic. Therefore, no finger 
tip-splitting is observed even at third order. 

A possible exception to this rule arises from a nonlinear instability
caused by the coupling of the harmonic mode to itself. Inspecting
equation~(\ref{single}) we note that high wavenumber modes with $k >
\sqrt{2/5B}$ (the harmonic mode satisfies this condition) are unstable
for sufficiently large amplitude, because the coefficient of
$\zeta_k^3$ is positive. This large-amplitude instability was
previously noted in the related solidification problem by Dee and
Mathur~\cite{Mathur}. Finger tip-splitting in the Saffman-Taylor
problem, induced by large amplitude noise assisted by a nonlinear
instability, has been suggested on the basis of numerical
simulations~\cite{DeG,Mei} and seen in
experiments~\cite{Par2,Tab,Max,Arn}, performed at very low $B$.

\section{Concluding remarks}
In this paper we developed a mode coupling theory to investigate the 
onset of nonlinear effects in the viscous fingering problem in 
a rectangular Hele-Shaw cell. 
From a weakly nonlinear analysis of the system, 
we derived a mode coupling differential equation which describes the 
evolution of the interface perturbation amplitudes. 
The basic symmetries of the mode coupling equation are identified and 
discussed. 

We investigated the relation between the viscosity contrast $A$ and
the interfacial asymmetry in the Saffman-Taylor problem. Viscosity
contrast $A$ multiplies symmetry-breaking terms in the mode coupling
equation. Our analysis explicitly indicates that symmetry breaking
occurs through enhanced growth of sub-harmonic perturbations. We show
that second-order terms that drive tip-splitting in the radial flow
geometry are prohibited by symmetry, explaining the general absence of
tip-splitting in the rectangular geometry. A remaining mystery is to
explain why small wave number modes do not alter the growth of high
wave number modes at second order.

Our mode coupling analysis shows that modest growth of the harmonic
may occur at third order by a reduction of the effective linear
stability of the harmonic mode, coupled with a driving process
assisted by finger competition. We do not expect this process to
result in finger tip-splitting under normal circumstances. However,
self-coupling of the harmonic creates a nonlinear instability that
could split finger tips if noise or other factors drive a sufficiently
large harmonic amplitude. Finally, we identified the onset of
saturation effects, which moderates the exponential growth of the
linear instability.

In a separate work~\cite{jsp} we extend the present theory to a system
in which one of the two fluids is a ferrofluid~\cite{Ros}, and a
magnetic field is applied normal to the Hele-Shaw cell. Interfacial
symmetry breaking at late stages is very dramatic in this
system~\cite{Per}. We point out, here, that the onset of interface
symmetry breaking depends on viscosity contrast $A$, not on the
applied magnetic field. In reference~\cite{jsp} we show that finger
tip-splitting arises as a result of the application of an external
magnetic field.

\pagebreak

\noindent
{\bf Acknowledgments}\\
\noindent
J.A.M. (CNPq reference number 200204/93-9) 
would like to thank CNPq (Brazilian Research Council) for financial 
support. This work was supported in part by the National Science 
Foundation grant No. DMR-9221596.

\pagebreak

\noindent
{\Large {\bf Figure Captions}}
\vskip 0.5 in
\noindent
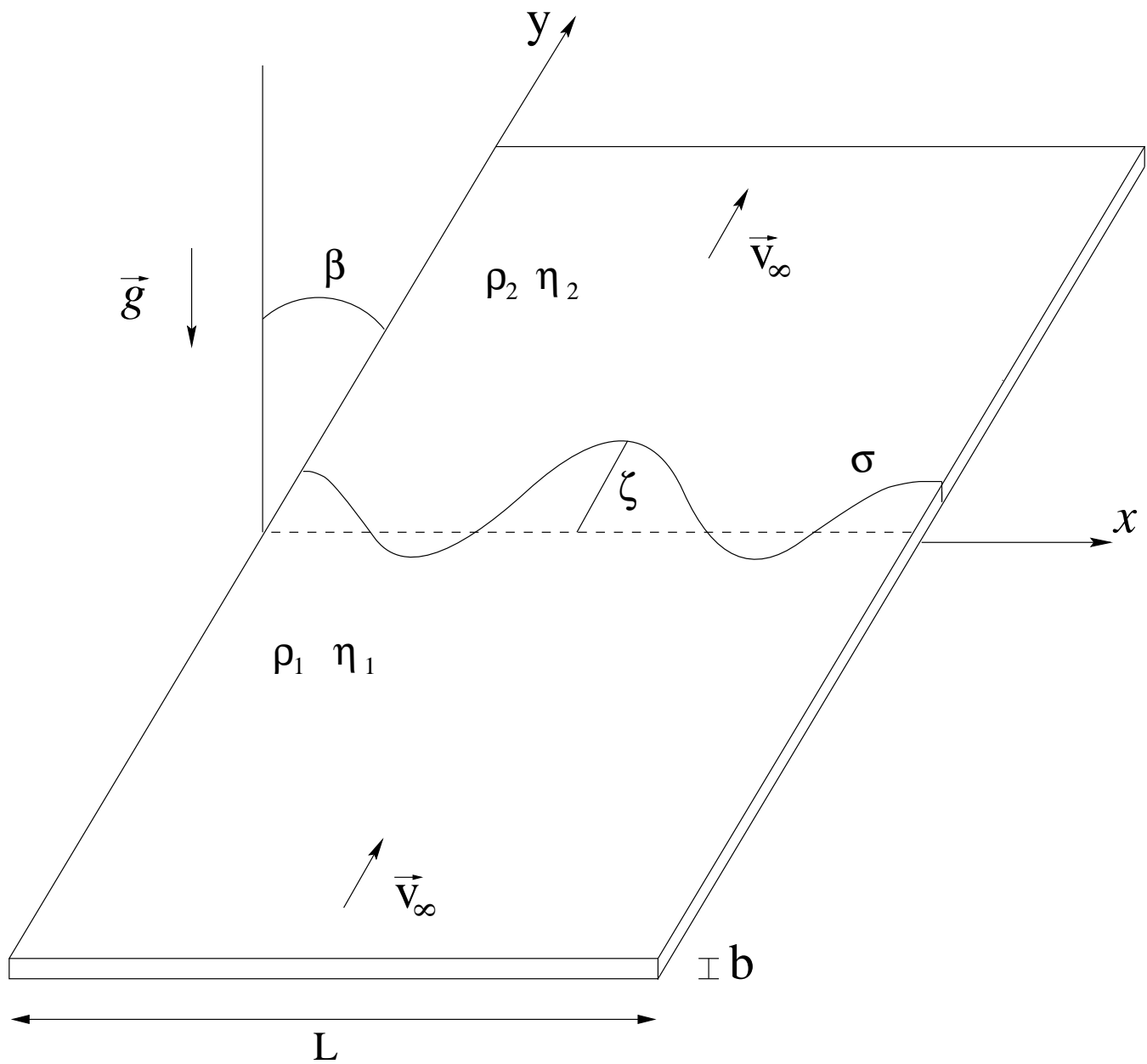
{\bf Figure 1:} Schematic configuration of the flow in a rectangular 
Hele-Shaw cell. The densities and viscosities of lower and upper fluids are 
$\rho_{1}$, $\eta_{1}$ and $\rho_{2}$, $\eta_{2}$, respectively. 
The dashed line represents the unperturbed interface $y=0$ and the solid 
undulated curve depicts the perturbed interface $y=\zeta(x,t)$, over 
the range $0 \le x \le L$.
The surface tension 
between the two fluids is given by $\sigma$ and the gravitational 
acceleration, directed in the negative vertical direction, 
is denoted by $\vec g$. The Hele-Shaw cell of thickness $b$ is tilted 
by an angle $\beta$ from the vertical direction. 
$\vec v_{\infty}=v_{\infty}\hat{y}$ represents the uniform overall velocity.
\vskip 0.25 in
\noindent
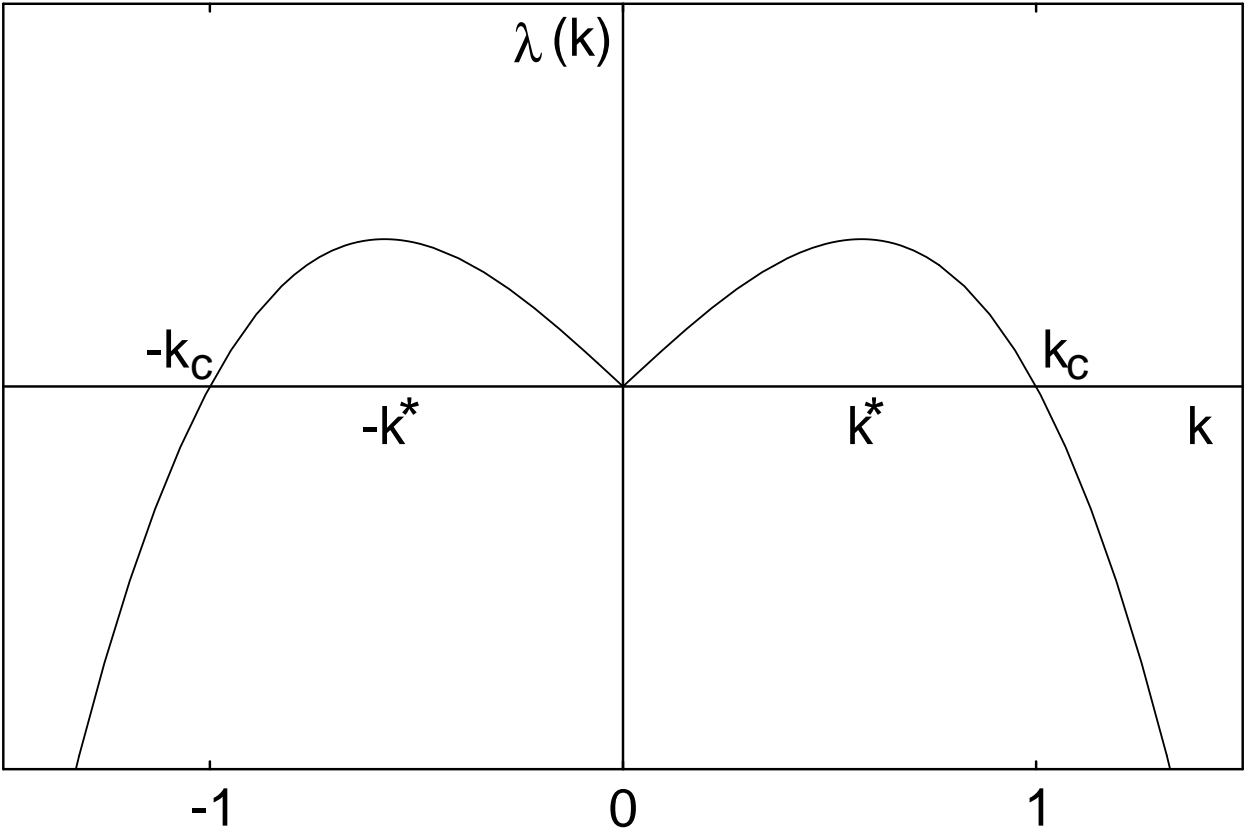
{\bf Figure 2:} Linear growth $\lambda(k)$ [see equation~(\ref{growth}) 
in the text] as a function of the wave number $k$, for $B=1$. The 
critical wave number $k_{c}=\pm 1$, and the most unstable wave number 
$k^{\ast}=\pm 1/\sqrt{3}$.
\vskip 0.25 in
\noindent
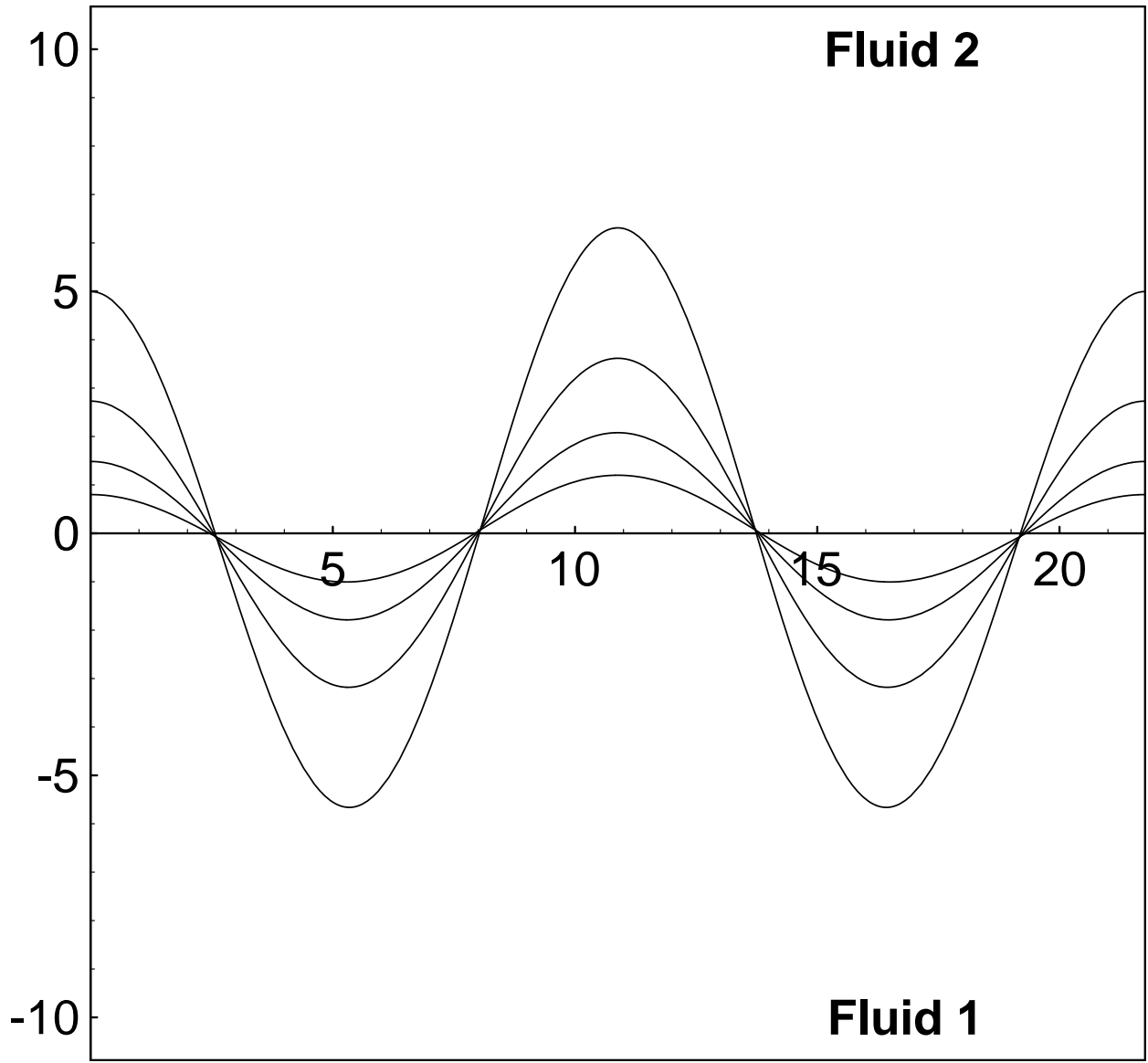
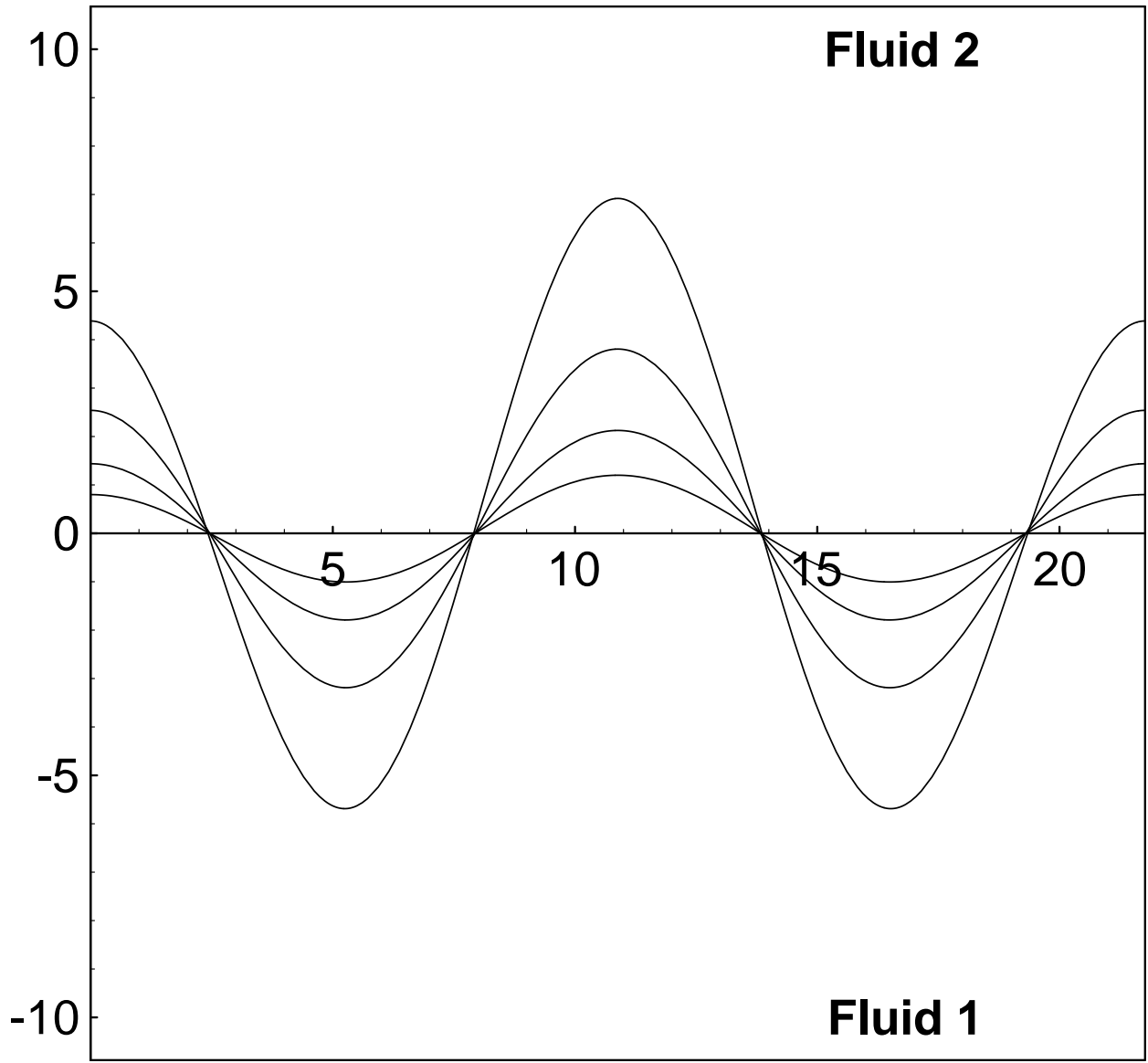
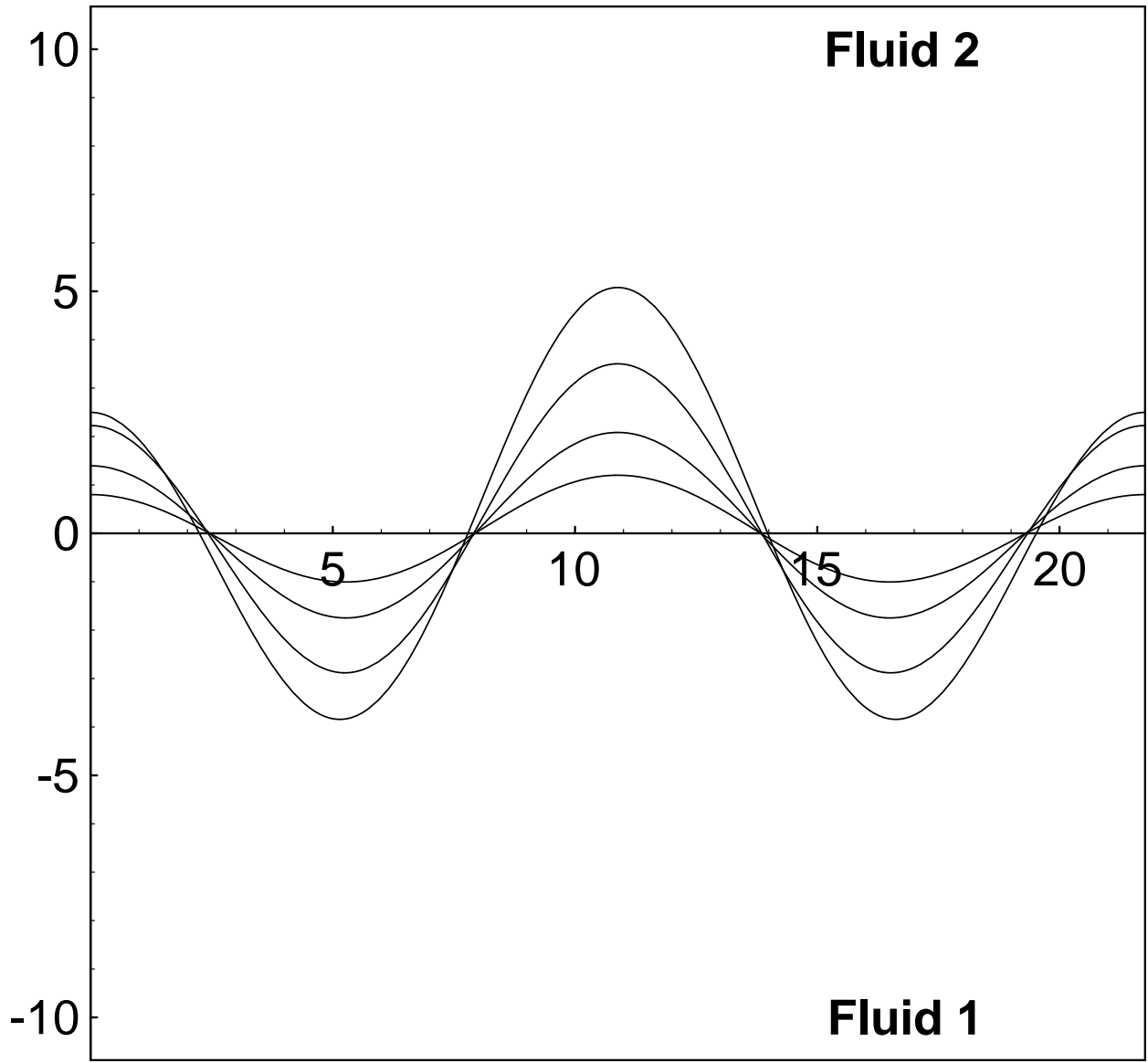
{\bf Figure 3:} Time evolution of the interface between the fluids, 
for the case of two interacting cosine waves. 
The dominant mode wave vector $k_{f}=k^{*}=1/\sqrt{3}$ has initial 
perturbation amplitude $a_{k_{f}}(0)$=$1$. 
The sub-harmonic mode wave vector $k_{s}=k_{f}/2$ and initial amplitude 
$a_{k_{s}}(0)=0.2a_{k_{f}}(0)$. 
For all panels the vertical axis scale is the same, 
the parameter $B=1$, and $t$=0, 1.5, 3.0 and 4.5; 
(a) First order (linear) solution; 
(b) Second order solution for $A=1$; 
(c) Third order solution for $A=1$. 
\vskip 0.25 in

\begin{thebibliography}{99}
\bibitem{Cap} The dimensionless surface tension parameter $B$ (defined in equation~(\ref{B})) is a measure of the surface tension forces relative to viscous forces. A low value of $B$ would mean, for instance, small surface tension or high speeds of flow.
\end{thebibliography}
\end{document}